\documentstyle[11pt,moriond,epsfig]{article}

\def\Journal#1#2#3#4{{#1} {\bf #2}, #3 (#4)}

\def\CPC{{\em Comp.\ Phys.\ Comm.}}
\def\EPJC{{\em Eur.\ Phys.\ Jour.} C}
\def\JHEP{{\em Jour.\ High En.\ Phys.}}

\def\NPB{{\em Nucl.\ Phys.} B}
\def\PLB{{\em Phys.\ Lett.}  B}

\def\PR{\em Phys.\ Rev.}
\def\PRD{{\em Phys.\ Rev.} D}
\def\ZPC{{\em Z. Phys.} C}

\def\be{\begin{equation}}
\def\ee{\end{equation}}
\def\bea{\begin{eqnarray}}
\def\eea{\end{eqnarray}}

\begin{document}
\vspace*{4cm}
\title{MULTIPLICITY RATIOS AND THE CONFINEMENT MECHANISM}

\author{ P. ED\'EN }

\address{Nordita, Blegdamsvej 17\\ DK-2100 Copenhagen O, Denmark}

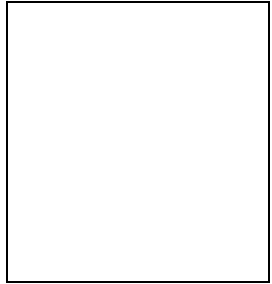
\maketitle\abstracts{
I review theoretical and experimental challenges in studies of the ratio of multiplicities in quark and gluon jets, and its relevance for our understanding of hadronization.}

\section*{Exordium}
At the best of conferences, a talk initiates reactions and discussions which are more relevant for the proceedings than the talk itself. New bright ideas initiated by the reactions could call for presentation, or there could be an apparent need to review the basic arguments, perused and put in perspective in the discussions. I have at this conference had the privilege of experiencing the latter. Therefore, this note is not so much about the latest calculations, presented here at Moriond 2001, but rather a review and summary of presentations during three years, triggered by the response at this conference.

\section*{Narratio}
The ratio of multiplicities in gluon and quark jets, $r$, at asymptotic energies, is predicted by perturbative QCD (pQCD) since long.\cite{RG,MLLA}  Renormalization group (RG) calculations~\cite{RG} and the modified leading log approximation (MLLA)~\cite{MLLA} of parton showers (PS) give identical results on the leading, ${\cal{O}}(\sqrt{\alpha_s})$, correction. At the higher of presently available energies, these  reduce the asymptotic value, 9/4, with about 10\%. The theoretical calculations apply to parton multiplicities, which are identified with hadron data using the assumption of Local Parton--Hadron Duality (LPHD).\cite{LPHD}

Monte Carlo simulations of PS give a result on $r$ which differs
significantly from the MLLA expectation, also at energies far
beyond reach today. Analytical investigations show
that energy conservation effects in the PS give rise to corrections which are
numerically large, though formally suppressed by a relative factor
$\alpha_s$.\cite{Dremin,g-jets} At $\mathrm{Z}$ energies, the PS ${\cal{O}}(\alpha_s)$ corrections exceeds the ${\cal{O}}(\sqrt{\alpha_s})$ ones.
Investigations of two alternative PS formulations, the parton-- and dipole cascade, respectively, give qualitatively similar results. A detailed comparison reveals some noticeable differences,\cite{compare} but much more striking is the difference between these PS results and the earlier RG calculation,\cite{RG} where ${\cal{O}}(\alpha_s)$ corrections are found to be negligible, typically of order 1\%.

\section*{Probatio}

In view of these differences, one should note that pure pQCD predictions can be obtained only for infrared safe quantities, for which the diverging density of soft partons is compensated by their vanishing influence on the observable. Multiplicities do not fall into this category of observables. Though there are ratios, like $r$, for which divergences cancel, the results depend strongly on assumptions about the relevance at hadron level of various soft partons. Thus, the huge difference between RG and PS results, and the sizeable difference between different PS versions, is not due to trivial errors, but may instead give insight to the confinement mechanism. 

Looking at the implicit hadronization assumption of the pQCD+LPHD approaches, one finds reason to take the PS results seriously. 
To avoid divergences, a PS calculations must be stopped at some infrared cut-off scale. When combined with the LPHD assumption, the cut-off quantity and value is assumed to properly omit partons irrelevant at hadron level. A PS with a locally invariant cut-off represents a hadronization model based on preconfinement,\cite{preconf} which implies that the effects of a parton with e.g.\ a red colour charge depends on where in phase space its partner antired charge is located. The parton and dipole cascades are two alternative formulations within this framework.

Today, two different hadronization models, cluster~\cite{herwig} and string~\cite{string} fragmentation, are generally successful in describing data. They are different in many respects, but preconfinement is a basic feature they have in common. 
The success of these phenomenological hadronization models based on preconfinement, in contrast to models with independent fragmentation, gives encouraging support to the PS approach.

More important support comes directly from data on $r$. PS results show good
simultaneous fit over scales ranging from $\Upsilon$ to Z energies,
both on $r$ and on the ratio of multiplicity slopes.\cite{CLEOBill,D2}

\section*{Refutatio}
With RG techniques, the ratio of parton multiplicities, in a fixed phase space region defining a jet, can be calculated. The preconfinement assumption, which claims that the relevance at hadron level of a parton depends on more local characteristics, is not represented by the RG approach. Nevertheless it is a very rigorous calculation and the firmly established belief that ``NNLO corrections always are small, (if you know where to look)''~\cite{Yuri} calls for further defense of the large ${\cal{O}}(\alpha_s)$ PS corrections. It seems, therefore, appropriate to comment on some possible objections, and proposed alternatives, to the PS results.

The first impression of accumulated experimental reports on the subject may be that there is little room for firm conclusions. This, however, stems from complications in extracting good gluon jet data. To understand the problem, consider two-jet events in $e^+e^-$ annihilation, selected with some jet clustering algorithm. While the multiplicity in the full event sample depends on the event energy only, the sub-sample is biased by the jet resolution scale. The multiplicity in these biased events depends on both energy and jet resolution, which cannot be combined into one single effective scale. In general, the theoretically considered jets correspond to a hemisphere in an unbiased event sample, while most jet clustering algorithms define jets corresponding to a hemisphere of an imagined two-jet event.
Some optimistic attempts have been made to characterize conventional jets with one single scale,\cite{badscale,D1} but multiplicity studies more closely connected to theoretical predictions are also available.\cite{CLEOBill,D2,g-jets,3-jet} These show up a consistent picture, supporting the PS results on multiplicities.

Non-perturbative corrections of  different origin than the PS energy conservation effect are in general more heavily suppressed. Experience is~\cite{D1} that a tuning of these, without the PS ${\cal{O}}(\alpha_s)$ terms,  to the large corrections needed at Z energies give unreasonable predictions at the $\Upsilon$ scale. Furthermore, the heavily suppressed corrections tend to have minor effect on the ratio of slopes. Just as for $r$, this quantity can be reliably confronted with data only after careful gluon jet definitions. The latest \textsc{Delphi} result~\cite{D2} agrees with the PS picture, and if other LEP1 experiments were to follow with similar analyses, there is hope for a more conclusive combination of results.

The PS energy conservation effect is numerically large, and can not be compensated by other ${\cal{O}}(\alpha_s)$ corrections. Though the PS is known to differ from correct many-parton emission densities with terms of relative order $\alpha_s$, the corrections can be effectively (and efficiently) included with a finite change of $\Lambda_{\mathrm {QCD}}$.\cite{Lambda} The resulting ``bremsstrahlung $\Lambda$'' is less than a factor 2 larger than $\Lambda_{\mathrm {\overline{MS}}}$. In Monte Carlo simulation programs, $\Lambda$ is treated as a free parameter,\cite{herwig,ariadne} in general tuned to a value in reasonable agreement with the bremsstrahlung $\Lambda$. Numerically, a change of $\Lambda$ within reasonable limits (say, a factor 4) give effects far too small to influence the discussion and conclusion in this note.

The preconfinement assumption relies on a matching of colour-- and anticolour charges into singlet systems. Due to the finite number of colours, this matching is not unique, but the resulting possibility of so called colour reconnection has not been manifested anywhere yet, in spite of some impressing and dedicated analyses.\cite{CR} Observation of colour reconnection would have fundamental influence on our understanding of confinement, but its numerical effect on inclusive multiplicity ratios is minor.

\section*{Peroratio}
Ratios of multiplicities are sensitive to hadronization, in a way quite different from most kinematic infrared safe observables. Due to the lack of purely gluonic sources at high energies, great care is needed in the design of experimental analyses. As that care is now beginning to be taken, theorists should be open-minded and acknowledge the phenomenological uncertainties related to their predictions. If so, these relatively simple observables can be a great probe of the confinement mechanism.

\section*{Acknowledgments}
I thank Yuri Dokshitzer for stimulating discussions and G\"osta Gustafson, Valery Khoze  and Igor Dremin for fruitful cooperations.

\end{document}